\documentclass[aps,prl,superscriptaddress,twocolumn]{revtex4} 
\usepackage{graphicx}
\usepackage{epstopdf}
\usepackage{bm}
\usepackage{xcolor}

\begin{document}

\newcommand{\etal}{{\it et al.}\/}
\newcommand{\gtwid}{\mathrel{\raise.3ex\hbox{$>$\kern-.75em\lower1ex\hbox{$\sim$}}}}
\newcommand{\ltwid}{\mathrel{\raise.3ex\hbox{$<$\kern-.75em\lower1ex\hbox{$\sim$}}}}

\title{Effective pairing interaction in a system with an incipient band}

\author{T.~A.~Maier}
\affiliation{Computational Sciences and Engineering Division and Center for
Nanophase Materials Sciences, Oak Ridge National Laboratory, Oak Ridge,
Tennessee 37831-6164, USA}

\author{V.~Mishra}
\affiliation{Computational Sciences and Engineering Division and Center for
Nanophase Materials Sciences, Oak Ridge National Laboratory, Oak Ridge,
Tennessee 37831-6164, USA}

\author{D.~J.~Scalapino}
\affiliation{Department of Physics, University of California, Santa Barbara,
CA 93106-9530, USA}

\date{\today}

\begin{abstract} The nature and mechanism of superconductivity in the
extremely electron-doped FeSe based superconductors continues to be a matter
of debate. In these systems, the hole-like band has moved below the Fermi
energy, and various spin-fluctuation theories involving pairing between states
near the electron Fermi surface and states of this incipient band have been
proposed. Here, using a dynamic cluster quantum Monte Carlo calculation for a
bilayer Hubbard model we show that the pairing in these systems can be
understood in terms of an effective retarded attractive interaction between
electrons near the electron Fermi surface. \end{abstract}


\maketitle



The proposal that spin-fluctuation scattering of pairs between the electron
and hole Fermi surfaces of the Fe-based superconductors provides the pairing
mechanism in these materials is challenged \cite{ref:1} by the occurrence of
superconductivity in FeSe monolayers on STO \cite{ref:2,ref:3,ref:4}, and K
and Li FeSe intercalates \cite{ref:5,ref:6,ref:7,ref:8,ref:9}. In these
materials the hole band near $\Gamma$ is submerged below the Fermi level
leaving only an electron-like Fermi surface (FS) around the M point. Various
authors have suggested that an $s^\pm$ pairing state can be formed in which a
gap appears on the incipient band with the opposite sign to the gap on the
electron Fermi surface \cite{ref:10,ref:11,ref:12}.

Here using a dynamic cluster approximation (DCA) \cite{MaierRMP05} quantum
Monte Carlo (QMC) calculation for a bilayer Hubbard model we show that this
physics can be described in terms of an effective pairing interaction for the
fermions near the electron Fermi surface. Unlike the usual momentum
dependent spin-fluctuation pairing interaction, this effective interaction is
essentially independent of momentum transfer, but depends upon the Matsubara
frequency transfer. It is local in space but retarded in time. While the
resulting superconducting state is similar to that of the incipient band
pairing proposals, the introduction of an effective interaction provides a
different perspective on the pairing interaction. In this case, just as in the
traditional electron-phonon superconductors, it is the frequency dependence of
the pairing interaction rather than its momentum dependence that is important.
As a consequence, it is the sign change of the gap with frequency that
characterizes the pairing.

The system we will study is a bilayer Hubbard model with
\begin{eqnarray}          
	H&=&t\sum_{\langle ij\rangle
	m\sigma}(c^\dagger_{jm}c^{\phantom\dagger}_{im}+{\rm h.c.})+t_\perp
\sum_{i\sigma}(c^\dagger_{i1\sigma}c^{\phantom\dagger}_{i2\sigma}+{\rm
h.c.}) \nonumber \\          &&-\mu\sum_{im\sigma}n_{im\sigma}+U\sum_{
im}n_{im\uparrow}n_{im\downarrow}\,. \label{eq:1} 
\end{eqnarray} 
The operator $c^\dagger_{im\sigma}$ creates an electron on the $i^{\rm th}$
site of the $m=1$ or 2 layer with spin $\sigma$ and
$n_{im\sigma}=c^\dagger_{im\sigma}c^{\phantom\dagger}_{im\sigma}$. Here $t$ is
the intra-layer near neighbor hopping, $t_\perp$ the inter-layer hopping and
$U$ the on site interband Coulomb interaction. The bandstructure for periodic
boundary conditions is
\begin{equation}
\varepsilon_{\bm k}=2t(\cos k_x+\cos k_y)+t_\perp\cos k_z-\mu\,. \label{eq:2}
\end{equation}

The results which will be shown are obtained from a DCA calculation on a
$(4\times4)\times 2$ cluster with 16 sites in each layer. In the DCA, the
momentum space is coarse-grained and thereby the lattice problem is reduced to
a finite size cluster embedded in a mean-field that is self-consistently
determined to represent the remaining lattice degrees of freedom
\cite{MaierRMP05}. This 32-site cluster problem is then solved with a
continuous-time auxiliary-field QMC algorithm \cite{GullEPL08,GullPRB11}.
While the ${\bm k}$ dependence of irreducible quantitites, i.e. the
single-particle self-energy and the irreducible two-particle vertex functions,
is reduced to the 32 cluster momenta, the full lattice ${\bm k}$-dependence is
retained in the Green's function through the dispersion Eq.~(\ref{eq:2})
\cite{MaierRMP05}. In this approximation, correlations within a length-scale
set by the cluster size are treated accurately by QMC, while longer-ranged
correlations beyond the cluster are treated at a mean-field level.

\begin{figure*}[htbp]
\includegraphics[width=\textwidth]{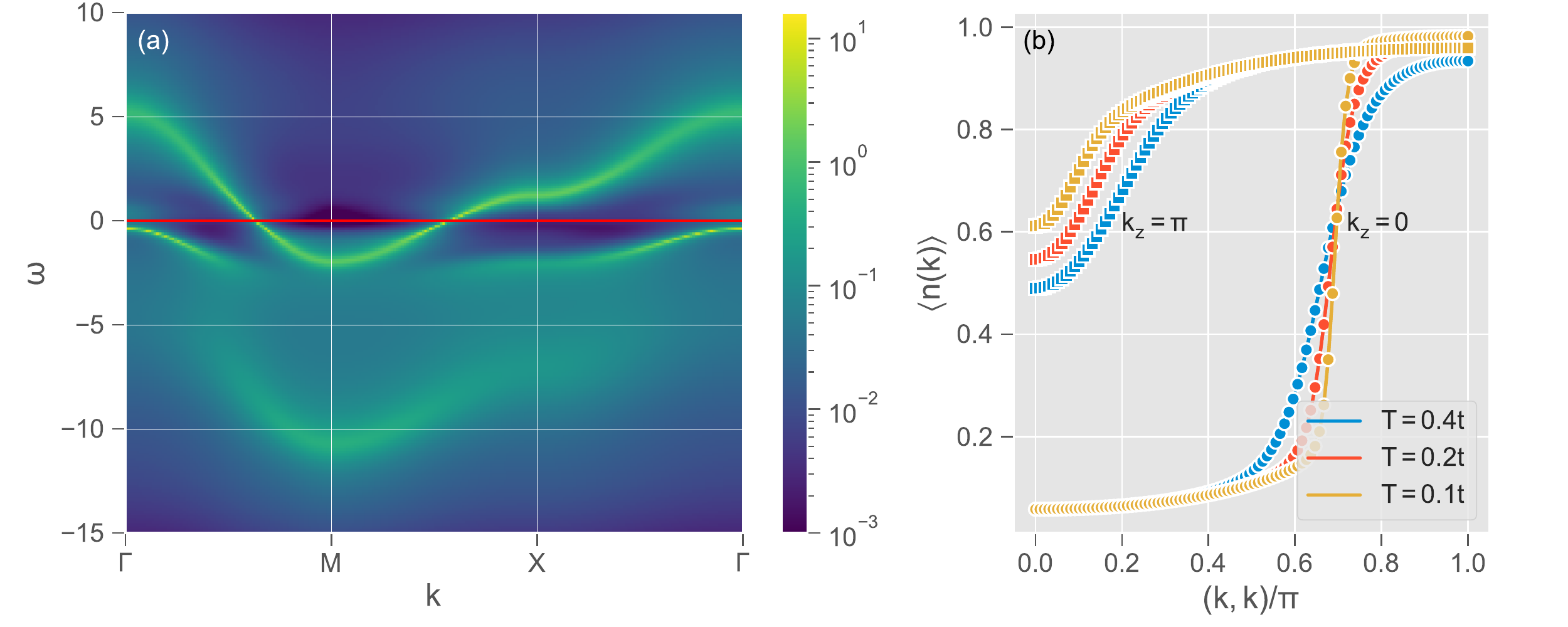}
  \caption{(a) The single particle spectral weight $A({\bm k},\omega)$ at
	$T=0.2$ for $t_\perp=2.5$, $n=1.15$ and $U=8$ shows an electron pocket
	around the $(k_x=\pi,\ k_y=\pi)$ $M$ point for $k_z=0$ and an
	incipient hole band laying below the Fermi energy and centered at the
	$(k_x=0,\ k_y=0)$ $\Gamma$ point for $k_z=\pi$. (b) The momentum
	occupation $\langle n({\bm k})\rangle$ plotted for ${\bm k}$ running
	from $\Gamma$ to M sharpens at the electron FS as the temperature
	is decreased, and is suppressed at $\Gamma$ as the $k_z=\pi$ band
	drops below the Fermi energy.
	\label{fig:1}}
\end{figure*}

\begin{figure}[htbp]
\includegraphics[width=0.48\textwidth]{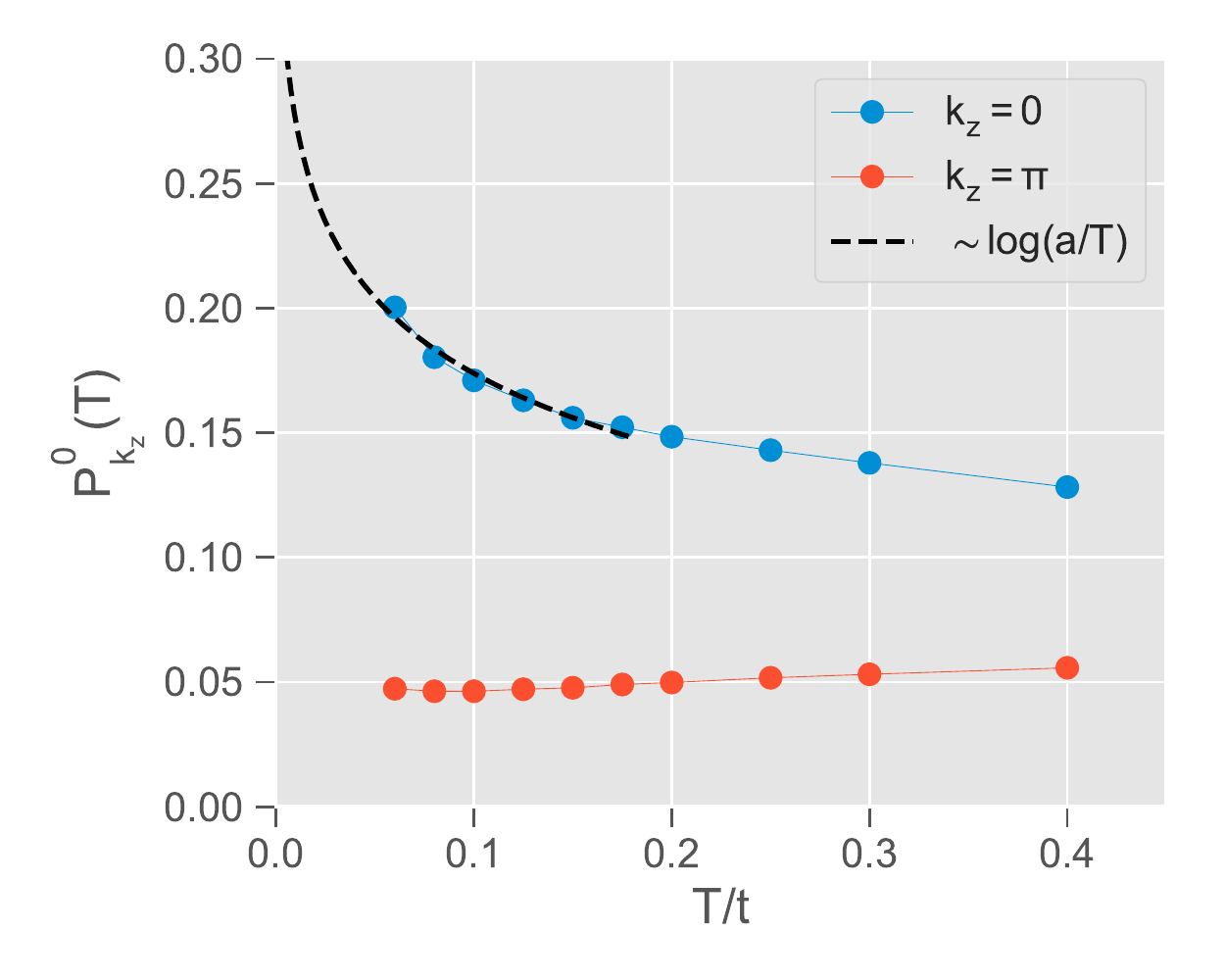}
  \caption{The intrinsic pairfield susceptibilities $P^0_0(T)$ and $P^0_\pi(T)$
	versus $T$. $P^0_0(T)$ exhibits a $\log(t/T)$ Cooper instability
	associated with the $k_z=0$ Fermi surface while $P^0_\pi(T)$ for the
	incipient band remains flat as $T$ decreases.
	\label{fig:2}}
\end{figure}

In the following we will set $t_\perp/t=2.5$, $U/t=8$ and take a filling
$n=1.15$. For these parameters, the single-particle spectral weight $A({\bm
k},\omega)$, obtained from a Maximum Entropy estimation, plotted in
Fig.~\ref{fig:1}(a),
shows evidence of a $k_z=0$ electron Fermi surface around the $(k_x=\pi,\
k_y=\pi)$ $M$ point and an incipient $k_z=\pi$ hole band that has dropped just
below the Fermi energy at the $\Gamma$ point. Consistent with this, the
momentum distribution $\langle n({\bm k})\rangle$ for $k_z=0$ shown in
Fig.~\ref{fig:1}(b) sharpens at ${\bm k}_F$ as the temperature is reduced,
while for $k_z=\pi$, $\langle n({\bm k})\rangle$ fades away as $T$ decreases
indicating that the $k_z=\pi$ band lays below the Fermi energy. Further
evidence of an incipient $k_z=\pi$ band is seen in the behavior of the
intrinsic pairfield susceptibility
\begin{equation}
  P^0_{k_z}(T)=\frac{T}{N}\sum_{k,\omega_n}\phi^2(\omega_n)G_{k_z}({\bm
  k},\omega_n)G_{k_z}(-{\bm k},-\omega_n)
\label{eq:3}
\end{equation}
Here we have used a smooth frequency cut-off $\phi(\omega_n) = (\pi^2 T^2 +
\omega_c^2) / (\omega_n^2+\omega_c^2)$ with $\omega_c=t$. $G_{k_z}({\bm
k},\omega_n)$ is the dressed single particle propagator associated with the
$k_z=0$ or $\pi$ bands, ${\bm k}=(k_x,k_y)$ and $\omega_n=(2n+1)\pi T$ is a
Matsubara frequency. The intrinsic pairfield susceptibility is expected to
exhibit a Cooper $\log(t/T)$ behavior when there is a Fermi surface. As shown
in Fig.~\ref{fig:2},
one sees this type of behavior for the $k_z=0$ electrons, but $P^0_\pi(T)$ for
the $k_z=\pi$ hole band remains flat as $T$ decreases.

\begin{figure*}[t]
\includegraphics[width=0.9\textwidth]{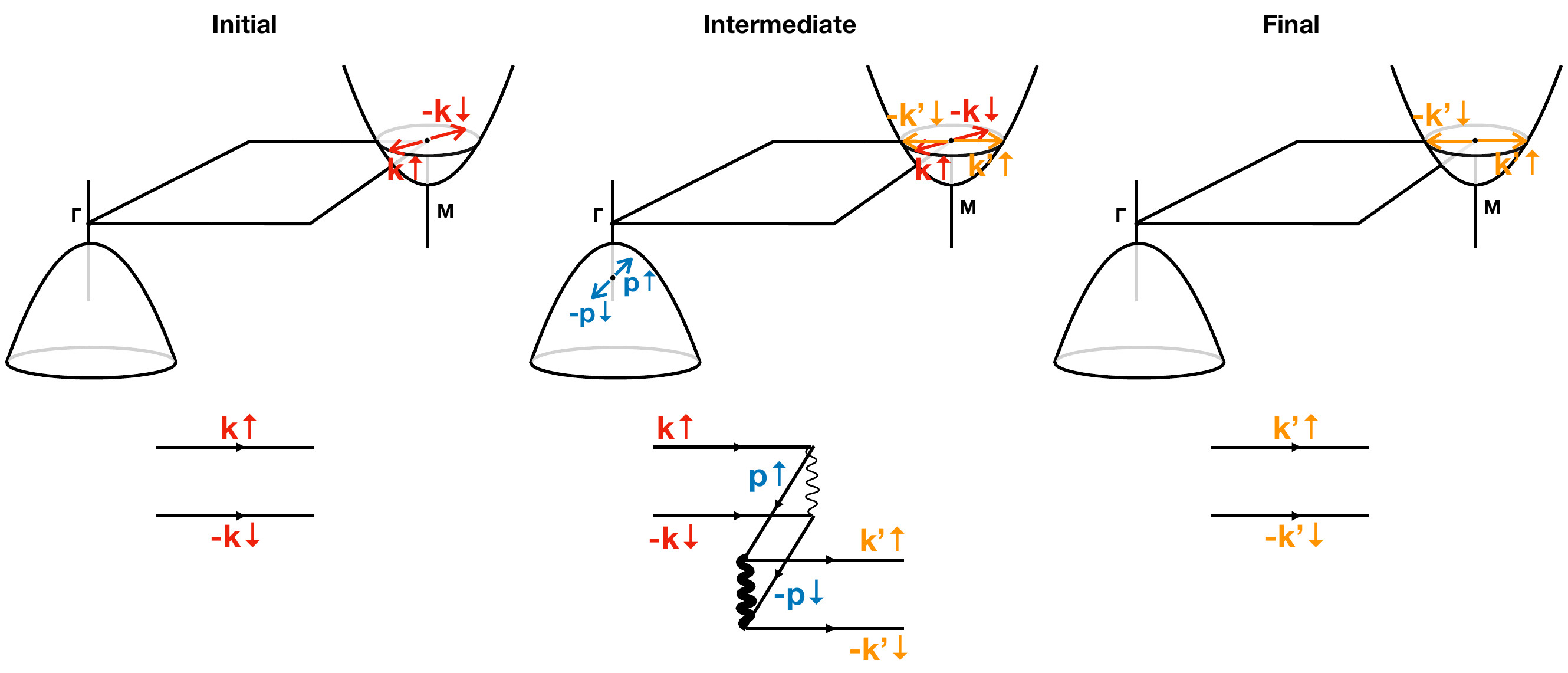}
  \caption{Schematic illustration of the effective intermediate state
	contribution to the $\Gamma_2$ pairing vertex. The Feynman diagram for
	$\Gamma_2$ shown in the lower part of the figure involves the intermediate
	states illustrated above it. The thin interaction line denotes the
	$\Gamma_{0\pi}^0$ fully irreducible vertex and the thicker interaction
	line the $\Gamma_{\pi 0}$ vertex, which is only irreducible in the $k_z=0$
	two-particle channel.
	\label{fig:3}}
\end{figure*}

With this in mind, we consider the pairing interaction $\Gamma(K,K')$ with
$K=(k_x,k_y,i\omega_n)$ and $K'=(k'_x,k'_y,i\omega_{n'})$ between fermion
pairs near the $k_z=0$ electron FS. This interaction can be separated into two
particle-particle scattering vertices
\begin{equation}
  \Gamma(K,K')=\Gamma_1(K,K')+\Gamma_2(K,K')\,.
\label{eq:4}
\end{equation}
The first of these, $\Gamma_1$, involves intermediate pair scattering
processes near the electron Fermi surface $(k_z=0)$ while $\Gamma_2(K,K')$
involves the incipient $(k_z=\pi)$ hole band. A schematic illustration of
$\Gamma_2(K,K')$ is shown in Fig.~\ref{fig:3}.
\begin{figure*}[]
\includegraphics[width=\textwidth]{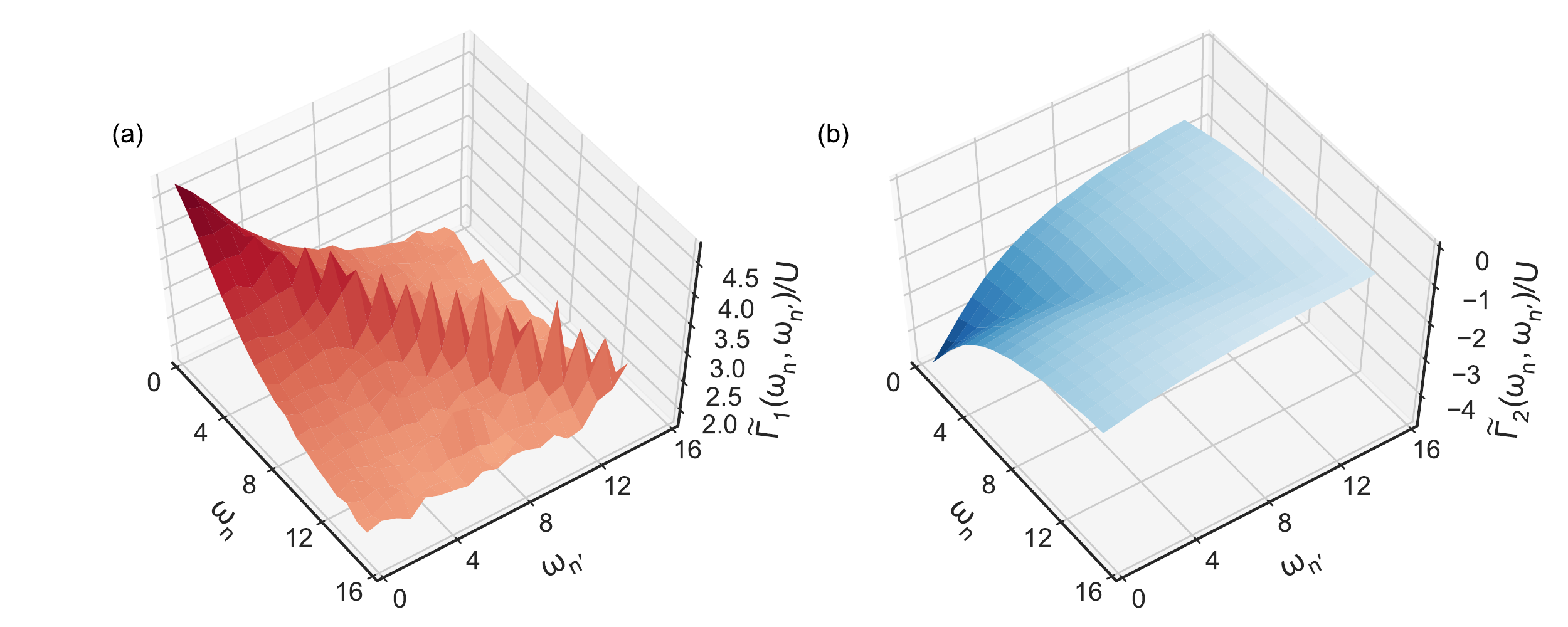}
  \caption{The two particle scattering vertices for $T=0.125t$, normalized to
	$U$, versus the Matsubara energies $\omega_n$ and $\omega_{n'}$. (a)
	$\tilde\Gamma_1(\omega_n,\omega_{n'})$ and (b)
	$\tilde\Gamma_2(\omega_n,\omega_{n'})$.
	\label{fig:4}}
\end{figure*}

$\Gamma_1$ is irreducible in both the $k_z=0$ and $k_z=\pi$ particle-particle
channels while $\Gamma_2$ is irreducible in only the $k_z=0$ channel. As shown
in the Feynman diagram in Fig.~\ref{fig:3}, $\Gamma_2$ can be written as
\begin{eqnarray}
  \Gamma_2(K,K')=-\frac{T}{N}\sum_{K''}&&\hspace{-0.25cm}\Gamma^0_{0\pi}(K,K'')G_\pi(K'')G_\pi(-K'')\nonumber\\
                                       &&\hspace{-0.25cm}\times\,\,\Gamma_{\pi0}(K'',K')\,.
\label{eq:5}
\end{eqnarray}
Here the $\Gamma^0_{0\pi}$ vertex involves pairs with $k_z=0$ near the
electron FS which scatters to states in the incipient $k_z=\pi$ band. It is
irreducible in both the $k_z=0$ and $\pi$ two particle channels. The single
particle Green's function $G_{\pi}(K'')$ is the dressed electron propagator on
the incipient band and the vertex $\Gamma_{\pi0}$ is only irreducible in the
$k_z=0$ channel, so that $\Gamma_2$ contains multiple scattering processes
involving pairs on the incipient $k_z=\pi$ band.

The important momentum dependence of the pairing interaction which involves
the inter-band $k_z-k'_z=\pi$ scattering has been separated out and the
$\Gamma$ vertices in Eq.~(\ref{eq:5}) are slowly varying functions of
$k_x-k'_x$ and $k_y-k'_y$. The important variables are the Matsubara energies
$\omega_n$ and $\omega_{n'}$. The gap is an even function of $\omega_n$ so
that it is useful for plotting to introduce symmetrized vertices \cite{ref:13}
\begin{equation}
  \tilde\Gamma(\omega_n,\omega_{n'})=\Gamma(\omega_n,\omega_{n'})+\Gamma(\omega_n,-\omega_{n'})
\label{eq:6}
\end{equation}
with $\omega_n$ and $\omega_{n'}$ varying over positive Matsubara frequencies.
Results for $\tilde\Gamma_1(\omega_n,\omega_{n'})$ and
$\tilde\Gamma_2(\omega_n,\omega_{n'})$ are plotted in Fig.~\ref{fig:4}~(a-b)
for ${\bm k}$ and ${\bm k}'$ set to $(\pi,\pi)$.  The contribution to the
pairing interaction from  pair scatterings on the FS,
$\tilde\Gamma_1(\omega_n,\omega_{n'})$, is positive while the contribution
from the virtual pair scattering involving the $k_z=\pi$ band,
$\tilde\Gamma_2(\omega_n,\omega_{n'})$, is negative. The strength of the
attractive $\tilde\Gamma_2$ is associated with the spin-fluctuation
$k'_z-k_z=\pi$ scattering processes that scatter pairs between the electron
Fermi surface and the incipient band. It is this transfer rather than the
scattering interactions on the incipient band that is important. The
$\omega_{n'}=\omega_n$ cleft in $\tilde\Gamma_2(\omega_n,\omega_{n'})$ shown
in  Fig.~\ref{fig:4}~(b) corresponds to having zero center of mass energy in
this transfer process.

Combining $\Gamma_1$ and $\Gamma_2$, the resulting effective pairing interaction
$\Gamma$ is attractive at low Matsubara frequencies and repulsive at higher
frequencies.
\begin{figure}[htbp]
\includegraphics[width=0.45\textwidth]{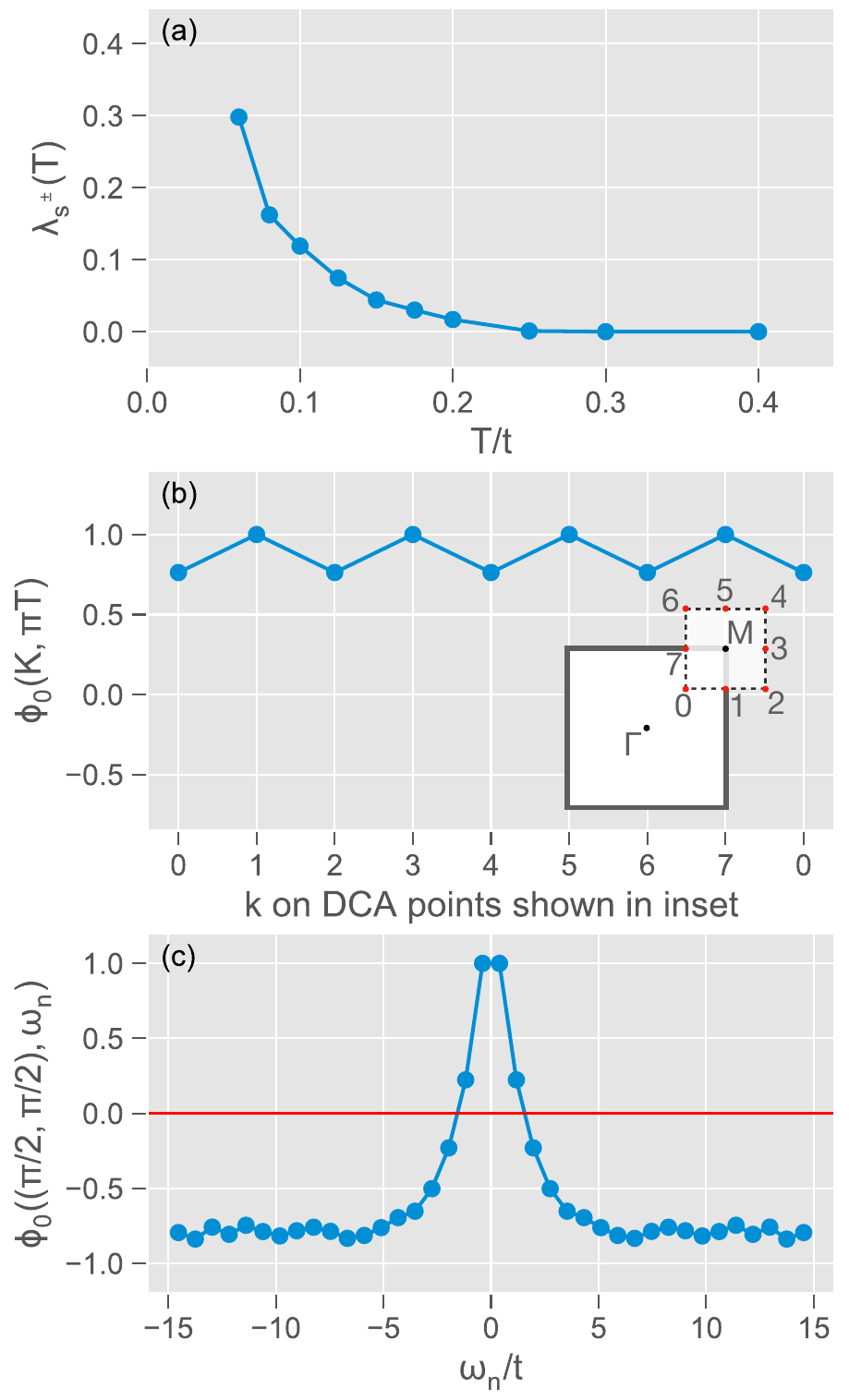}
  \caption{(a) The leading eigenvalue $\lambda_s(T)$ of the Bethe-Salpeter
	equation as $T$ decreases. (b) The momentum dependence of the gap
function $\phi({\bm k},\omega_n)$ of the leading eigenvalue for $\omega_0=\pi
T$ for ${\bm k}$ values near the FS as shown in the inset and (c) its
frequency dependence for ${\bm k}=(\pi/2,\pi/2)$ for $T=0.125t$. $\phi({\bm
k},\omega_n)$ changes sign leading to a reduction of the effect of the
repulsive $\Gamma_1$ interaction on $\lambda_s(T)$.
	\label{fig:5}}
\end{figure}
Using $\Gamma$ in the Bethe-Salpeter equation \cite{ref:13}
\begin{equation}
  -\frac{T}{N}\sum_{K'}\Gamma(K,K')G_\pi(K')G_\pi(-K')\phi(K')=\lambda\phi(K)
\label{eq:6}
\end{equation}
with $K=(k_x,k_y,\omega_n)$, we find the leading eigenvalue shown in
Fig.~\ref{fig:5}a and the ${\bm k}$ and $\omega_n$ dependence of the
eigenfunction $\phi({\bm k},\omega_n)$ shown in 5b and 5c. As noted, the
irreducible vertex $\Gamma(K,K')$, and therefore the eigenfunction $\phi(K)$
only depend on the 32 DCA cluster momenta, while the Green's function $G(K)$
retains the full momentum dependence of the lattice \cite{MaierPRL06}. The
eigenfunction $\phi({\bm k},\omega_n)$ 
is essentially independent of ${\bm k}$ as shown in  Fig.~\ref{fig:5}b but
changes sign as $\omega_n$ increases. This sign change is such that the gap
function is positive over the frequency regime characteristic of the
spin-fluctuations. This is similar to the frequency dependence of the gap in
the traditional electron-phonon-Coulomb problem. The eigenfunction $\phi({\bm
k},\omega_n)$ is positive when $|\omega_n - \omega_{n'}|$ is less than several
times the energy of the spin-fluctuation exchange and then changes sign at
higher frequencies leading to a suppression of the repulsive part of the
potential. Putting this another way, if the interaction were cut-off when
$|\omega_n - \omega_{n'}|$ exceeded several times the exchange energy, the
remaining part of $\Gamma_1$ would be replaced by a smaller pseudopotential.

To summarize, in this picture (1) antiferromagnetic order is suppressed as the
hole (or electron) band becomes incipient, leaving strong $k'_z-k_z = \pi$
spin-fluctuations and (2) the pairing interaction arises from these 
spin fluctuation scattering processes which involve  intermediate states on
the incipient band and give rise to an attractive retarded pairing interaction
for the fermions on the remaining electron Fermi surface.

\section*{Acknowledgments}
This work was supported by the Scientific Discovery through Advanced Computing
(SciDAC) program funded by the U.S. Department of Energy, Office of Science,
Advanced Scientific Computing Research and Basic Energy Sciences, Division of
Materials Sciences and Engineering. An award of computer time was provided by
the INCITE program. This research used resources of the Oak Ridge Leadership
Computing Facility, which is a DOE Office of Science User Facility supported
under Contract DE-AC05-00OR22725.




\begin{thebibliography}{99}
\bibitem{ref:13} While it is convenient to plot the $\omega_n$ and $\omega_
	{n'}$ dependence of the symmetrized vertex $\tilde\Gamma$, as shown in
	Fig.~\ref{fig:4}, in the Bethe-Salpeter equation (\protect\ref{eq:6}) we
	have used the unsymmetrized vertex so that in principle one could have
	odd-frequency $p$-wave singlet and $s$-wave triplet solutions. We find
	that the even frequency $s$-wave singlet has the leading eigenvalue for
	temperatures $T\lesssim 0.15t$.
\end{thebibliography}
\end{document}